\documentclass[12pt]{article}
\usepackage{epsf}
\usepackage{amsmath}
\usepackage{graphics}
\usepackage{cite}

\renewcommand{\thefootnote}{\#\arabic{footnote}}
\begin{document}

\newcommand{\gtrsim}{ \mathop{}_{\textstyle \sim}^{\textstyle >} }
\newcommand{\lesssim}{ \mathop{}_{\textstyle \sim}^{\textstyle <} }

\newcommand{\rem}[1]{{\bf #1}}

\renewcommand{\thefootnote}{\fnsymbol{footnote}}
\setcounter{footnote}{0}
\begin{titlepage}

\def\thefootnote{\fnsymbol{footnote}}

\begin{center}
\hfill hep-th/0603065\\
\hfill March 2006\\
\vskip .5in
\bigskip
\bigskip
{\Large \bf Anomaly Cancellation and Conformality in Quiver Gauge Theories}

\vskip .45in

{\bf Edoardo Di Napoli and Paul H. Frampton}

{\em Perimeter Institute, 31 Caroline Street, Waterloo, ON N2L 2Y5, Canada}\\

and

{\em University of North Carolina, Chapel Hill, NC 27599-3255, USA \footnote{Permanent address}}

\end{center}

\vskip .4in
\begin{abstract}

Abelian quiver gauge theories provide nonsupersymmetric candidates for
the conformality approach to physics beyond the standard model. Written
as ${\cal N}=0$, $U(N)^n$ gauge theories, however, they have 
mixed $U(1)_p U(1)_q^2$ and $U(1)_p SU(N)_q^2$
triangle anomalies. It is shown how to construct 
explicitly a compensatory term $\Delta{\cal L}_{comp}$
which restores gauge invariance of ${\cal L}_{eff} = {\cal L} + \Delta {\cal L}_{comp}$
under $U(N)^n$. 
It can lead to a negative contribution to the U(1) $\beta$-function and
hence to one-loop conformality at high energy for all
dimensionless couplings.
\end{abstract}
\end{titlepage}

\renewcommand{\thepage}{\arabic{page}}
\setcounter{page}{1}
\renewcommand{\thefootnote}{\#\arabic{footnote}}
\newpage

\bigskip

\noindent {\it Introduction}

\bigskip

One alternative to supersymmetry and grand unification is to postulate
conformality, four-dimensional conformal invariance at high energy, for
the non gravitational extension of the standard model. Although much
less vigorously studied than supersymmetry, the conformality
approach suggested\cite{PHF1998} in 1998 has made considerable progress. 
Models which contain the standard model fields have been constructed\cite{Z7}
and a model which grand unifies at about 4 TeV\cite{Z12} has been
examined.

Such models are inspired by the AdS/CFT correspondence\cite{Maldacena}
specifically based on compactification of the IIB superstring
on the abelian orbifold $AdS_5 \times S^5/Z_n$ with N coalescing
parallel D3 branes. A model is specified by N and by the embedding
$Z_n \subset SU(4)$ which is characterized by integers
$A_m$ ($m = 1,2,3,4$) which specify how the {\bf 4} of $SU(4)$
transforms under $Z_n$.
Only three of the $A_m$ are independent because
of the $SU(4)$ requirement that $\Sigma_m A_m = 0$ (mod n).
The number of vanishing $A_m$ is the
number ${\cal N}$ of surviving supersymmetries. Here we
focus on the non supersymmetric ${\cal N}=0$
case.

In \cite{CFR}, the original speculation \cite{PHF1998}
that such models may be conformal has been refined to
exclude models which contain scalar fields
transforming as adjoint representations because only if all
scalars are in bifundamentals are there chiral fermions and,
also only if all scalars are in bifundamentals, the
one-loop quadratic divergences cancel in the scalar propagator.
We regard it as encouraging that these two desirable
properties select the same subset of models.

Another phenomenological encouragement stems from the
observation\cite{Vafa} that the standard model representations for
the chiral fermions can all be accommodated in bifundamentals
of $SU(3)^3$ and can appear naturally in the conformality
approach. 

In the present article we address the issue of triangle anomalies.
Although the purely non abelian anomalies involving $SU(N)^3$
subgroups of the $U(N)^n$ gauge group are cancelled, 
there do survive triangle anomalies
of the types $U(1)_p U(1)_q^2$ and $U(1)_p SU(N)_q^2$. 
Since the original superstring
is anomaly free, one expects such anomalies to be cancelled. This
cancellation is well understood\cite{stringtheory}
in terms of the closed string axions
coupling to $F \tilde{F}$. Here we shall 
construct a compensatory term $\Delta {\cal L}_{comp}$ which
is non polynomial in the bifundamental scalars and which
when added to the gauge lagrangian ${\cal L}$ gives
rise to an effective lagrangian ${\cal L}_{eff}
= {\cal L} + \Delta {\cal L}_{comp}$ which is $U(N)^n$
gauge invariant.

In the next subsection, we shall discuss the anomaly
cancellation by a compensatory term. The following
section explains the explicit construction. There is
then a treatment of the evolution of the U(1) couplings
and finally there is some discussion.

\newpage

\bigskip

\noindent {\it Anomaly cancellation by a compensatory term}

\bigskip

The lagrangian for the nonsupersymmetric $Z_n$ theory can be written in
a convenient
notation which accommodates simultaneously both adjoint and
bifundamental scalars as
\begin{eqnarray}
{\cal L} & = &
-\frac{1}{4} F_{\mu\nu; r,r}^{ab}F_{\mu\nu; r,r}^{ba}
+i \bar{\lambda}_{r + A_4, r}^{ab} \gamma^{\mu} D_{\mu} \lambda_{r,
r+A_4}^{ba} \nonumber \\
& &
+ 2 D_{\mu} \Phi_{r+a_i, r}^{ab \dagger} D_{\mu} \Phi_{r, r+a_i}^{ba}
+i \bar{\Psi}_{r+A_m, r}^{ab} \gamma^{\mu} D_{\mu} \Psi_{r, r+A_m}^{ba}
\nonumber \\
&  &
- 2 i g
\left[ \bar{\Psi}_{r, r+A_i}^{ab} P_L \lambda_{r + A_i, r + A_i +
A_4}^{bc}
\Phi_{r + A_i+A_4, r}^{\dagger ca}
- \bar{\Psi}_{r, r+A_i}^{ab} P_L \Phi_{r + A_i, r - A_4}^{\dagger bc}
\lambda_{r - A_4, r}^{ca}
\right] \nonumber \\
& & -   \sqrt{2} i g \epsilon_{ijk}
\left[
\bar{\Psi}_{r, r + A_i}^{ab} P_L \Psi_{r +A_i, r + A_i + A_j}^{bc}
\Phi_{r -A_k - A_4, r}^{ca}
-
\bar{\Psi}_{r, r + A_i}^{ab} P_L \Phi_{r +A_i, r + A_i + A_k +
A_4}^{bc} \Psi_{r - A_j, r}^{ca}
\right] \nonumber \\
& & - g^2 \left(
\Phi_{r, r + a_i}^{ab} \Phi_{r+a_i,r}^{\dagger bc}
-
\Phi_{r, r - a_i}^{\dagger ab} \Phi_{r-a_i,r}^{bc}
\right)
\left(
\Phi_{r, r + a_j}^{cd} \Phi_{r+a_j,r}^{\dagger da}
-
\Phi_{r, r - a_j}^{\dagger cd} \Phi_{r- a_j,r}^{da}
\right)  \nonumber \\
& & + 4 g^2
\left(
\Phi_{r, r+a_i}^{ab}\Phi_{r+a_i, r+a_i+a_j}^{bc}
\Phi_{r+a_i+a_j,r+a_j}^{\dagger cd}\Phi_{r+a_j,r}^{\dagger da}
\right.   \nonumber \\
&  &
-
\left. \Phi_{r, r+a_i}^{ab}\Phi_{r+a_i, r+a_i+a_j}^{bc}
\Phi_{r+a_i+a_j,r+a_i}^{\dagger cd}\Phi_{r+a_i, r}^{\dagger da}
\right)
\label{N=0L}
\end{eqnarray}
where $\mu, \nu = 0, 1, 2, 3$ are lorentz indices; $a, b, c, d = 1$ to
$N$ are $U(N)^n$
group labels; $r = 1$ to $n$ labels the node of the quiver diagram
(when the two node subscripts are equal it is an adjoint plus singlet
and the two superscripts are in the same U(N): when the two node
subscripts are unequal it is a bifundamental and the two superscript
labels transform under different U(N) groups);
$a_i ~~ (i = \{1, 2, 3\}) $ label the first three of the {\bf 6} of
SU(4);
$A_m ~~ (m = \{1, 2, 3 ,4\}) = (A_i, A_4)$ label the {\bf 4} of SU(4).
By definition
$A_4$ denotes an arbitrarily-chosen fermion ($\lambda$) associated with
the gauge boson,
similarly to the notation in the ${\cal N} = 1$ supersymmetric case.
Recall that $\sum_{m=1}^{m=4} A_m = 0 $ (mod n).

As mentioned above we shall restrict attention to models
where all scalars are in bifundamentals which
requires all $a_i$ to be non zero. Recall that
$a_1=A_2+A_3$, $a_2=A_3+A_1$; $a_3=A_1+A_2$.

The lagrangian in Eq(\ref{N=0L}) is classically $U(N)^p$ gauge
invariant. There are, however, triangle anomalies of the
$U(1)_p U(1)^2_q$ and $U(1)_p SU(N)_q^2$ types. Making gauge transformations under
the $U(1)_r$ (r = 1,2,...,n) with gauge parameters $\Lambda_r$
leads to a variation

\begin{equation}
\delta {\cal L} = - \frac{g^2}{4\pi^2}\Sigma_{p=1}^{p=n} A_{pq} F_{\mu\nu}^{(p)}
\tilde{F}^{(p) \mu\nu} \Lambda_q
\label{Apq}
\end{equation}
which defines an $n \times n$ matrix $A_{pq}$ which is given by

\begin{equation}
A_{pq} = {\rm Tr} (Q_p Q_q^2)
\label{chiraltrace}
\end{equation}
where the trace is over all chiral fermion links and $Q_r$ is the
charge of the bifundamental under $U(1)_r$. We shall adopt the sign
convention that ${\bf N}$ has $Q=+1$ and ${\bf N^{*}}$ has $Q=-1$.

It is straightforward to write $A_{pq}$ in terms of
Kronecker deltas because the content of chiral fermions is

\begin{equation}
\Sigma_{m=1}^{m=4} \Sigma_{r=1}^{r=n} ({\bf N}_r, {\bf N^{*}}_{r+A_{m}})
\label{fermions}
\end{equation}
This implies that the antisymmetric matrix $A_{pq}$ is explicitly

\begin{equation}
A_{pq} = - A_{qp} = \Sigma_{m=1}^{m=4} \left( \delta_{p, q-A_{m}} -
\delta_{p, q+A_{m}} \right)
\label{ApqKronecker}
\end{equation}

\bigskip

Now we are ready to construct ${\cal L}_{comp}^{(1)}$, the compensatory
term. Under the $U(1)_r$ gauge tansformations with
gauge parameters $\Lambda_r$ we require that

\begin{eqnarray}
\delta {\cal L}_{comp}^{(1)} & = &   - \delta {\cal L} \nonumber \\
& = & + \frac{g^2}{4\pi^2}\Sigma_{p=1}^{p=n} A_{pq} F_{\mu\nu}^{(p)}
\tilde{F}^{(p) \mu\nu} \Lambda_q
\label{compensatory}
\end{eqnarray}
To accomplish this property, we construct a compensatory term in the form
\footnote{For a related construction in a different context, see\cite{Dudas}}
\begin{equation}
{\cal L}_{comp}^{(1)} = \frac{g^2}{4 \pi} \Sigma_{p=1}^{p=n}
\Sigma_{k} B_{pk} {\rm Im} {\rm Tr} {\rm ln}
\left( \frac{\Phi_k}{v} \right) F_{\mu\nu}^{(p)} \tilde{F}^{(p) \mu\nu}
\label{compensatory2}
\end{equation}
where $\Sigma_{k}$ runs over scalar links.
We believe this {\it form} for the compensatory term
to be unique\footnote{Although the general form is unique, there can be a technical
ambiguity in the matrix $B$ to be discussed below.} because 
${\cal L}_{comp}^{(1)}$ must be invariant under $SU(N)^n$. To see
that ${\cal L}_{comr}^{(1)}$ of Eq.(\ref{compensatory2}) has such invariance
rewrite Tr ln $\equiv$ exp det and note that the $SU(N)$
matrices have unit determinant. It is inconceivable that any other
non-trivial function of the bifundamental, other than a closed
loop of links in the quiver diagram, has the full
$SU(N)^n$ invariance but a closed loop, unlike Eq.(\ref{compensatory2}),
is $U(N)^n$ invariant.

We note {\it en passant} that one cannot take the 
$v \rightarrow 0$ limit in Eq.(\ref{compensatory2}); the chiral anomaly
enforces a breaking of conformal invariance.

Explicit construction of the matrix $B_{pk}$ will the subject of the subsequent
section. But first we investigate the transformation properties
of the ``$Im Tr ln (\Phi/v)$" term in Eq.(\ref{compensatory2}).
Define a matrix $C_{kq}$ by

\begin{equation}
\delta \left( \Sigma_{p=1}^{p=n} \Sigma_k {\rm Im}
{\rm Tr} {\rm ln} \left( \frac{\Phi_k}{v} \right) \right)
= \Sigma_{q=1}^{q=n} C_{kq} \Lambda_q
\label{Ckq}
\end{equation}
whereupon Eq.(\ref{compensatory}) will be satisfied if
the matrix $B_{pk}$ satisfies $A=BC$. The inversion $B=AC^{-1}$ is 
non trivial because $C$ is singular but $C_{kq}$ can be written in terms
of Kronecker deltas by noting that the content of
complex scalar fields in the model is

\begin{equation}
\Sigma_{i=1}^{i=3} \Sigma_{r=1}^{r=n} \left( {\bf N}_r, {\bf N^{*}}_{r \pm a_{i}}
\right))
\label{scalars}
\end{equation}
which implies that the matrix $C_{kq}$ must be of the form
\begin{equation}
C_{kq} = 3 \delta_{kq} - \Sigma_{i} \delta_{k+a_{i},q}
\label{CkqKronecker}
\end{equation}

\newpage

\bigskip

The $U(1)_p SU(N)_q^2$ triangle anomalies necessitate the addition
of a second compensatory term ${\cal L}_{comp}^{(2)}$. The derivation
of ${\cal L}_{comp}^{(2)}$ is similar to, but algebraically simpler than,
that for ${\cal L}_{comp}^{(1)}$. Under $U(1)_r$ with gauge parameter
$\Lambda_r$ and $SU(N)_s$ gauge transformations
the variation in ${\cal L}$ of Eq.(\ref{N=0L}) is
\begin{equation}
\delta {\cal L} = - \frac{g^2}{4\pi^2}\Sigma_{p=1}^{p=n} A_{pq}i^{'} 
F_{\mu\nu \alpha_p}^{(p) \beta_p}
\tilde{F}^{(p) \mu\nu \alpha_p}_{\beta_p} 
\Lambda_q
\label{A'pq}
\end{equation}
which defines an $n \times n$ matrix $A^{'}_{pq}$ which is given by

\begin{equation}
A_{pq}^{'} = {\rm Tr} (Q_p n_q) 
\label{chiraltraceprime}
\end{equation}
where the trace is over all chiral fermion links, $Q_r$ is the
charge of the bifundamental under $U(1)_r$ and 
$n_q$ is the number of fundamentals and anti fundamentals
of $SU(N)_q$ corresponding to all fermionic links
between nodes $p$ and $q$. As before, we adopt the sign
convention that ${\bf N}$ has $Q=+1$ and ${\bf N^{*}}$ has $Q=-1$.

It is straightforward to write $A_{pq}^{'}$ in terms of
Kronecker deltas  as

\begin{equation}
A_{pq}^{'} = - A_{qp}^{'} = \Sigma_{m=1}^{m=4} \left( - \delta_{p, q-A_{m}}  +
\delta_{p, q+A_{m}} \right)
\label{ApqKroneckerprime}
\end{equation}

\bigskip

\noindent ${\cal L}_{comp}^{(2)}$, the compensatory
term for the $U(1)_p SU(N)_q^2$ triangle anomalies is
\begin{equation}
{\cal L}_{comp}^{(2)} = \frac{g^2}{4 \pi} \Sigma_{p=1}^{p=n}
\Sigma_{k} B_{pk}^{'} {\rm Im} {\rm Tr} {\rm ln}
\left( \frac{\Phi_k}{v} \right) 
F_{\mu\nu \alpha_p}^{(p) \beta_p}
\tilde{F}^{(p) \mu\nu \alpha_p}_{\beta_p} 
\label{compensatoryprime2}
\end{equation}
where $\Sigma_{k}$ runs over scalar links.

Explicit construction of the matrix $B_{pk}^{'}$ in ${\cal L}_{comp}^{(2)}$
is more straightforward 
than $B_{pk}$ in ${\cal L}^{(1)}$ because when we define a matrix $C_{kq}^{'}$ by
the variation under mixed abelian-nonabelian gauge transformations
\begin{equation}
\delta \left( \Sigma_{p=1}^{p=n} \Sigma_k {\rm Im}
{\rm Tr} {\rm ln} \left( \frac{\Phi_k}{v} \right) \right)
= \Sigma_{q=1}^{q=n} C_{kq}^{'} \Lambda_q
\label{Ckqprime}
\end{equation}
we find $C_{kq}^{'} = 3\delta_{pk}$ so $B_{pk}^{'}$ in Eq.(\ref{compensatoryprime2}) is
$B_{pq}^{'} = \frac{1}{3} A_{pq}^{'}$ with $A_{pq}^{'}$ defined by Eq.
(\ref{ApqKroneckerprime})

\bigskip

\noindent {\it Explicit construction of the matrix B in ${\cal L}_{comp}^{(1)}$}

\bigskip

Construction of the anomaly compensatory term ${\cal L}_{comp}^{(1)}$
of Eq.(\ref{compensatory2}) has been reduced to the explicit
construction of the matrix $B_{pk}$. Although $B = A C^{-1}$ is inadequate because
Rank(C) $<$ n, a necessary and sufficient condition for the
existence of $B$ is Rank (A) $\leq$ Rank (C). Proving this
in general would be one approach but the large number of
special cases will make the proof lengthy. 
Of course, we strongly suspect that
the matrix $B$ must exist from indirect
string theory arguments\cite{stringtheory}
but we shall convince the reader directly by explicit
construction of B in two extremes which we call the totally
degenerate and the totally nondegenerate cases respectively.

Given the form of $A_{pq}$ in Eq.(\ref{ApqKronecker}) and of $C_{kq}$
in Eq.(\ref{CkqKronecker}), it is irresistible to make a corresponding
ansatz for $B_{pk}$

\begin{equation}
B_{pk} = \Sigma_{\eta} C_{\eta} \delta_{p,k+\eta}
\label{Bansatz}
\end{equation}
and this ansatz works by setting up recursion relations
for the $C_{\eta}$ and allows explicit solution for
the matrix B in {\it any} special case. Writing a general formula
for B will now be demonstrated in two extreme cases.

\bigskip

\noindent
{\tt Totally degenerate case}

\bigskip

\noindent We assume $A_m = (A, A, A, -3A)$ (modulo n). In this
case, from Eq.(\ref{ApqKronecker}),

\begin{equation}
A_{pq} = 3 \delta_{p,q-A} - 3\delta_{p,q+A} + \delta_{p,q+3A}
-\delta_{p,q-3A}
\label{Amatrix}
\end{equation}
and from Eq.(\ref{CkqKronecker})
\begin{equation}
C_{kq} = 3 (\delta_{kq} - \delta_{k+2A,q})
\label{Cmatrix}
\end{equation}
Using Eq.(\ref{Bansatz}) and comparing coefficients gives
the series of recursion relations
\begin{eqnarray}
3C_{-A} - 3C_{A} & = & 3 \\
3C_{A} - 3 C_{3A} & = & -3 \\
3C_{-3A} - 3C_{-A} & = & -1 \\
3C_{3A} - 3C_{5A} & = & +1 \\
3C_{-5A} - 3C_{-3A} & = & 0 \\
3C_{5A} - 3C_{7A} & = & 0
\end{eqnarray}
and so on, with solution $C_A = -2/3, C_{-A} = C_{3A} = 1/3$ and all other $C_A = 0$.
The explicit B matrix is thus

\begin{equation}
B_{pk} = \frac{1}{3} \left( -2\delta_{p,k+A} + \delta_{p,k+3A}
+\delta_{p,k-A} \right)
\label{Bmatrix}
\end{equation}
From Eqs.(\ref{Amatrix},\ref{Bmatrix},\ref{Cmatrix}) one confirms A = BC.

\newpage

\bigskip

\noindent {\tt Totally nondegenerate case}

\bigskip

At an opposite extreme we may assume that

\begin{equation}
\pm A_m ~~ {\rm and} ~~ (a_i \pm A_m) ~~ {\rm are ~~ all~~ nondegenerate ~~ integers ~~ (modulo ~~ n)}
\label{nondegenerate}
\end{equation}

\bigskip

\noindent Assumption (\ref{nondegenerate}) requires $n \gg 1$ and
so is not a physical case. In this limit the recursion relations
become

\begin{eqnarray}
3C_{-A_m} - \Sigma_i C_{a_i - A_m} & = &  +1 \\
3C_{A_m} - \Sigma_i C_{a_i + A_m} & = & -1
\label{nondegrec}
\end{eqnarray}

Because the $A_m$ enter symmetrically in the model, one can
put $C_{A_m} = x$ and $C_{-A_m}=y$ both independent of $m$. 
This yields $C_{a_i+A_m} = (x+\frac{1}{3})$ and $C_{a_i-A_m} = (y-\frac{1}{3})$.  
Comparing coefficients of Kronecker deltas gives $x=-2/9$
and $y=+1/9$ and hence an explicit form for $B_{pk}$ by substitution 
in Eq.(\ref{Bansatz}).

\bigskip

\noindent {\tt Intermediate cases}

\bigskip

When there are some degeneracies which violate assumption
(\ref{nondegenerate}), there are too many special cases to
permit any succinct general formula. Nevertheless, one can
find fairly easily the explicit B matrix for {\it any} specific model, as we
have done for dozens of cases, using Mathematica software.

\bigskip

\noindent {\tt Technical non uniqueness}

\bigskip

As mentioned in an earlier footnote, although the form
of ${\cal L}_{comp}^{(1)}$ in Eq.(\ref{compensatory2}) is unique
the matrix B can have technical non uniqueness which is best
illustrated by a specific example.

Take $n=6$ and the totally degenerate example $A_m = (1,1,1,3)$.
Following the analysis given earlier one finds for $C_{kq}$
\begin{equation}
C = \left( \begin{array}{cccccc}
3 & 0 & -3 & 0 & 0 & 0 \\
0 & 3 & 0 & -3 & 0 & 0 \\
0 & 0 & 3 & 0 & -3 & 0 \\
0 & 0 & 0 & 3 & 0 & -3 \\
-3 & 0 & 0 & 0 & 3 & 0 \\
0 & -3 & 0 & 0 & 0 & 3
\end{array}
\right)
\label{Cmatrix2}
\end{equation}
while for the matrix $B_{pk}$

\begin{equation}
3B = \left( \begin{array}{cccccc}
0 & 1 & 0 & 1 & 0 & -2 \\
-2 & 0 & 1 & 0 & 1 & 0 \\
0 & -2 & 0 & 1 & 0 & 1 \\
1 & 0 & -2 & 0 & 1 & 0 \\
0 & 1 & 0 & -2 & 0 & 1 \\
1 & 0 & 1 & 0 & -2 & 0 
\end{array}
\right)
\label{Bmatrix2}
\end{equation}

\newpage

\bigskip

Multiplication of BC gives the required

\begin{equation}
A = \left( \begin{array}{cccccc}
0 & 3 & 0 & 0 & 0 & -3 \\
-3 & 0 & 3 & 0 & 0 & 0 \\
0 & -3 & 0 & 3 & 0 & 0 \\
0 & 0 & -3 & 0 & 3 & 0 \\
0 & 0 & 0 & -3 & 0 & 3 \\
3 & 0 & 0 & 0 & -3 &0 
\end{array}
\right)
\label{Amatrix2}
\end{equation}

\bigskip

There is, however, the following matrix $\hat{B}$

\begin{equation}
\hat{B} = \left( \begin{array}{cccccc}
0 & 1 & 0 & 1 & 0 & 1 \\
1 & 0 & 1 & 0 & 1 & 0 \\
0 & 1 & 0 & 1 & 0 & 1 \\
1 & 0 & 1 & 0 & 1 & 0 \\
0 & 1 & 0 & 1 & 0 & 1 \\
1 & 0 & 1 & 0 & 1 &0 
\end{array}
\right)
\label{hatBmatrix}
\end{equation}
which has the property that $\hat{B}C = 0$.This means there
is a one parameter family $B^{'} = B + \alpha \hat{B}$
with $\alpha$ a continuous parameter which can be used
in ${\cal L}_{comp}$. However, this also means that
the term
\begin{equation}
\hat{{\cal L}} = \frac{g^2}{4 \pi} \Sigma_{p=1}^{p=n}
\Sigma_{k} \hat{B}_{pk} {\rm Im} {\rm Tr} {\rm ln}
\left( \frac{\Phi_k}{v} \right) F_{\mu\nu}^{(p)} \tilde{F}^{(p) \mu\nu}
\label{topological}
\end{equation}
is $U(N)^n$ invariant and therefore need not be added for
purposes of anomaly cancellation.

\bigskip
\bigskip

The non-uniqueness of the matrix $B$ under $B \rightarrow B + \alpha \hat{B}$
has no immediate physical interpretation but it may suggest an undiscovered residual
symmetry.

\newpage

\noindent {\it String theory}

\bigskip

In \cite{aldazabal} the anomaly $A_{pq}$ is written in a factorized form
$A=TU$ following from the closed string axion exchange using the
Green-Schwarz mechanism so here we 
compare the two factorized expressions $A=BC$ and $A=TU$ to
become convinced that there is no connection.

The expression from \cite{aldazabal} is
\begin{equation}
A_{pq} = \sum_{l=1}^{l=n} T_{pl}U_{lq}V_{l}
\label{A=TU}
\end{equation}
where
\begin{equation}
T_{pl} = {\rm exp}\left( \frac{2 \pi i pl}{n} \right) ~~~~
U_{lq} = {\rm exp}\left( \frac{- 2 \pi i lq}{n} \right) 
\label{TandU}
\end{equation}
and 
\begin{equation}
V_{l} = \Pi_{i=1}^{1=3} {\rm sin} \left( \frac{\pi l a_i}{n} \right)
\label{V}
\end{equation}

\bigskip

Let us take the example of $n=6$ and $A_m=(1, 1, 1, 3)$ for which the matrices
$A$, $B$ and $C$ are given above. In Eq.(\ref{A=TU}) there is an ambiguity
in whether $V$ is accommodated in $T$ or $U$ so let us look at three possibilities:
(i) $T^{'} = TV$  (ii) $U^{'}=VU$ and (iii) $T^{''}=T\sqrt{V}$, $U^{''}=\sqrt{V}U$.
The factorizing matrices are then, up to overall normalization which has no effect
on the matrix textures, as follows. We define $\alpha={\rm exp}(i \pi/3)$.
\begin{equation}
A = T^{'}U =
\left(
\begin{array}{cccccc}
 \alpha & \alpha^2 & 0 & \alpha & \alpha^2 &  0  \\
\alpha^2 & -\alpha & 0 & -\alpha^2 & \alpha & 0 \\
-1 & 1 & 0 & -1 & 1 & 0 \\
-\alpha & \alpha^2 & 0 & \alpha & -\alpha^2 & 0 \\
-\alpha^2 & -\alpha & 0 & -\alpha^2 & -\alpha & 0 \\
1 & 1 & 0 & -1 & -1 & 0
\end{array}
\right)
\left(
\begin{array}{cccccc}
 -\alpha^2 & -\alpha & -1 & \alpha^2 & \alpha &  1  \\
-\alpha & \alpha^2 & 1 & -\alpha & \alpha^2 & 1 \\
-1 & 1 & -1 & 1 & -1 & 1 \\
\alpha^2 & -\alpha & 1 & \alpha^2 & -\alpha & 1 \\
\alpha & \alpha^2 & -1 & -\alpha & -\alpha^2 & 1 \\
1 & 1 & 1 & 1 & 1 &1
\end{array}
\right)
\label{TU'}
\end{equation}

\begin{equation}
A = TU^{'} =
\left(
\begin{array}{cccccc}
 \alpha & \alpha^2 & -1 & -\alpha & -\alpha^2 &  1  \\
\alpha^2 & -\alpha & 1 & \alpha^2 & -\alpha & 1 \\
-1 & 1 & -1 & 1 & -1 & 1 \\
-\alpha & \alpha^2 & 1 & -\alpha & \alpha^2 & 1 \\
-\alpha^2 & -\alpha & -1 & \alpha^2 & \alpha & 1 \\
1 & 1 & 1 & 1 & 1 &1
\end{array}
\right)
\left(
\begin{array}{cccccc}
 -\alpha^2 & -\alpha & -1 & \alpha^2 & \alpha &  1  \\
-\alpha & \alpha^2 & 1 & -\alpha & \alpha^2 & 1 \\
0 & 0 & 0 & 0 & 0 & 0 \\
-\alpha^2 & \alpha & -1 & -\alpha^2 & \alpha & -1 \\
-\alpha & -\alpha^2 & 1 & \alpha & \alpha^2 & -1 \\
0 & 0 & 0 & 0 & 0 & 0
\end{array}
\right)
\label{T'U}
\end{equation}

\begin{equation}
A = T^{''}U^{''} =
\left(
\begin{array}{cccccc}
 \alpha & \alpha^2 & 0 & -i\alpha & -i\alpha^2 &  0  \\
\alpha^2 & -\alpha & 0 & i\alpha^2 & -i\alpha & 0 \\
-1 & 1 & 0 &  i & -i & 0 \\
-\alpha & \alpha^2 & 0 & -i\alpha & i\alpha^2 & 0 \\
-\alpha^2 & -\alpha & 0 & i\alpha^2 & i\alpha & 0 \\
1 & 1 & 0 & i & i & 0
\end{array}
\right)
\left(
\begin{array}{cccccc}
-\alpha^2 & -\alpha & -1 & \alpha^2 & \alpha &  1  \\
-\alpha & \alpha^2 & 1 & -\alpha & \alpha^2 & 1 \\
0 & 0 & 0 & 0 & 0 & 0 \\
i\alpha^2 & -i\alpha & i & i\alpha^2 & -i\alpha & i \\
i\alpha & i\alpha^2 & -i & -i\alpha & -i\alpha^2 & i \\
0 & 0 & 0 & 0 & 0 & 0
\end{array}
\right)
\label{T''U''}
\end{equation}

\bigskip

By comparing the matrix textures in
Eqs.(\ref{TU'},\ref{T'U},\ref{T''U''})
with those for matrices $C$ and $B$ in Eqs.(\ref{Cmatrix2},\ref{Bmatrix2})
we see that the factorization of the anomaly matrix is not simply
related. The factorization in
Eqs.(\ref{TU'},\ref{T'U},\ref{T''U''})
follows from the physical requirement of factorization at
the closed string axion pole in a string tree diagram.
There is no similar requirement that mandates our
factorization $A=BC$ but these matrices have
simple textures so the factorization is not surprising.

\bigskip
\bigskip

In particular, the field theoretical mechanism
of anomaly cancellation discussed here has no
connection to the string
theoretical Green-Schwarz mechanism.

\newpage

\noindent {\it Evolution of U(1) gauge couplings.}

\bigskip

In the absence of the compensatory term, the two independent $U(N)^n$
gauge couplings $g_N$ for SU(N) and $g_1$ for U(1) are taken to be
equal $g_N(\mu_0) = g_1(\mu_0)$ at a chosen scale, {\it e.g.} $\mu_0$=4 TeV \cite{Z12,CRT}, 
to enable cancellation of quadratic 
divergences\cite{CFR}. Note that the $n$ SU(N) couplings 
$g_N^{(p)}$ are equal by the overall $Z_n$ symmetry, 
as are the $n$ U(1) couplings $g_1^{(p)}$, $1 \le p \le n$.

As one evolves to higher scales $\mu > \mu_0$, the renormalization
group beta function $\beta_N$ for SU(N)
vanishes $\beta_N =0$ at least at one-loop level so
the $g_N(\mu)$ can behave independent of the scale as expected by conformality.
On the other hand, the beta function $\beta_1$ for 
U(1)
is positive definite in the unadorned theory, given at one loop by, in the notation 
of \cite{GFT}
\begin{equation}
b_1 = \frac{11N}{48\pi^2}
\label{b1}
\end{equation}
where N is the number of colors. 
The corresponding coupling satisfies
\begin{equation}
\frac{1}{\alpha_1(\mu)} = \frac{1}{\alpha_1(M)} + 8\pi b_1 {\rm ln} \left( \frac{M}{\mu}
\right)
\end{equation}
so the Landau pole, putting $\alpha(\mu)=0.1$ and $N=3$, occurs at
\begin{equation}
\frac{M}{\mu} = {\rm exp} \left[ \frac{20 \pi}{11} \right] \simeq 302
\end{equation}
so for $\mu = 4$ TeV, $M \sim 1200$ TeV. The coupling becomes ``strong''
$\alpha(\mu) = 1$ at
\begin{equation}
\frac{M}{\mu} = {\rm exp} \left[ \frac{18 \pi}{11} \right] \simeq 171
\end{equation}
or $M \sim 680$ TeV.

We may therefore ask whether the new term ${\cal L}_{comp}$
in the lagrangian, necessary for anomaly cancellation, can
solve this problem for conformality?

Indeed there is the real counterpart of Eq,(\ref{compensatory2})
which has the form
\begin{equation}
{\cal L}_{comp}^{(1),real} = \frac{g^2}{4 \pi} \Sigma_{p=1}^{p=n}
\Sigma_{k} B_{pk} {\rm Re} {\rm Tr} {\rm ln}
\left( \frac{\Phi_k}{v} \right) F_{\mu\nu}^{(p)} F^{(p) \mu\nu}
\label{compensatoryreal}
\end{equation}
and this contributes to the U(1) gauge propagator
and to the U(1) $\beta-$function. Using
Eq.(\ref{Bmatrix}) for $B_{pk}$, the one-loop
quadratic divergence for a bifundamental scalar loop
cancels because
\begin{equation}
\Sigma_{k} B_{pk} = 0
\end{equation}
which confirms the cancellation found in \cite{CFR}.

Since the scale $v$ breaks conformal invariance,
the matter fields acquire mass, so the one-loop
diagram \footnote{The usual one-loop $\beta-$function is
of order $h^2$ regarded as an expansion in Planck's constant:
four propagators each $\sim h$ and two vertices each $\sim h^{-1}$
(c.f. Y. Nambu, Phys. Lett. {\bf B26,} 626 (1968)). 
The diagram considered
is also $\sim h^2$ since it has three propagators,
one quantum vertex $\sim h$ and an additional $h^{-2}$ associated with 
$\Delta m^2_{kk'}$.} has a logarithmic divergence proportional to

\begin{equation}
\int \frac{d^4p}{v^2} \left[ \frac{1}{(p^2-m_k^2)} - \frac{1}{(p^2-m_{k'}^2)}
\right]
\sim - \frac{ \Delta m^2_{kk'}}{v^2} {\rm ln} \left( \frac{\Lambda}{v} \right)
\end{equation}
the sign of which depends on $\delta m^2_{kk'} = (m_k^2 - m_{k'}^2)$.

To achieve conformality of U(1), a constraint must be imposed
on the mass spectrum of matter bifundamentals, {\it viz}
\begin{equation}
\Delta m^2_{kk'} \propto v^2 \left( \frac{11N}{48\pi^2} \right)
\end{equation}
with a proportionality constant of order one which depends on the choice
of model, the $n$ of $Z_n$ and the values chosen for $A_m, m=1,2,3$. This
signals how conformal invariance must be broken at
the TeV scale in order that it can be restored at high energy;
it is interesting that such a constraint arises 
in connection with an anomaly cancellation mechanism
which necessarily breaks conformal symmetry.

To give an explicit model, consider the case of $Z_4$
and $A_m = (1, 1, 1, 1)$ treated earlier for which 
one finds:
\begin{equation}
\Delta m^2_{kk'} = \frac{3}{2} v^2 \left( \frac{11N}{48\pi^2} \right)
\label{A111}
\end{equation}
In a more general model, the analog of Eq.(\ref{A111}) involves
replacement of $\frac{3}{2}$ by a generally different coefficient
derivable for each case from the coefficient $B_{pk}$ in 
Eq.(\ref{compensatory2}).

With such a constraint, the one-loop $\beta_1$ vanishes
in addition to $\beta_N$  so that the couplings $\alpha_1(\mu)$
and $\alpha_N(\mu)$ can be scale invariant for $\mu \ge \mu_0$.

For such conformal invariance at high energy to be maintained
to higher orders of perturbation theory
probably requires a global symmetry, for example the explicit form
of misaligned supersymmetry recently suggested in \cite{PHF2005}.

\newpage

\bigskip

\noindent {\it Discussion}

\bigskip

It has been shown how a compensatory term 
${\cal L}_{comp} = {\cal L}_{comp}^{(1)}
+ {\cal L}_{comp}^{(2)}$
can be constructed respectively to cancel the $U(1)_p U(1)_q^2$
and $U(1)_p SU(N)_q^2$
triangle anomalies in the quiver gauge theories with
chiral fermions.
We have emphasized the uniqueness of the
form of the compensatory term from the requirements
of invariance under $SU(N)^n \subset U(N)^n$.

Such a term can have phenomenological consequences. We
expect $v$ to be at the TeV scale as in \cite{Z12}
and ${\cal L}_{comp}$ reveals new non linear 
coupling between the bifundamental scalars and the
gauge fields expected to be significant in
the TeV energy regime.
Such empirical consequences merit further study.

It has further been shown that the compensatory
term ${\cal L}_{comp}$ can lead, with a suitable
mass spectrum of bifundamental matter, to vanishing one-loop
$\beta-$function for the U(1) gauge group, this raising the possibility
of one-loop scale invariance for all dimensionless couplings
which may persist at all higher loops in the presence
of a global symmetry.

\newpage

\begin{center}

{\bf Acknowledgements}

\end{center}

\bigskip
We wish to thank Freddy Cachazo for useful discussions
about string and field theories and their interconnection.
This work was supported in part by the
U.S. Department of Energy under Grant No. DE-FG02-97ER-41036.

\end{document}